\RequirePackage{fix-cm}
\documentclass[largeextended]{svjour3}       % onecolumn (second format)
\smartqed  % flush right qed marks, e.g. at end of proof
\usepackage{graphicx}
\usepackage{verbatim}  %add multiple lines of comments
\usepackage{marvosym}  %add the \Letter command
\usepackage{amsmath}
\usepackage{euscript}
\usepackage{wasysym}
\usepackage[section]{placeins}
 \usepackage{amsmath}
\usepackage{epstopdf}
\usepackage{amssymb}
\usepackage{tikz}
\usepackage{float}
\usepackage{diagbox}%for table
\usepackage{booktabs}%for table
\usepackage{mathdots}
\usepackage{mathrsfs}
\usepackage{caption}
\usepackage{algorithm}
\usepackage{algorithmic}
\usepackage{subfigure}
\usepackage{multirow}
\usepackage{array}
\usepackage{xhfill}
\usepackage{geometry}
\usepackage{subfig}
\usepackage{colortbl}
\usepackage{caption}
\usepackage{graphicx}
\usepackage{ulem}
\usepackage[justification=centering]{caption}% for centering the table caption
\usepackage{color}%
\usepackage{setspace}
\newcolumntype{A}{>{\columncolor{blue!50!white}}c}
\newcolumntype{B}{>{\columncolor{LightGoldenrod}}c}
\newcolumntype{C}{>{\columncolor{FireBrick!50}}c}
\newcolumntype{D}{>{\columncolor{Gray!42}}c}

\newcommand{\addfigure}[5][h]
{
	\begin{figure}[#1]
		\begin{center}
			\includegraphics[#5]{#2}
			\caption{#3.\label{#4}}
		\end{center}
	\end{figure}
}
\usepackage{geometry}
\begin{document}
\begin{spacing}{1.3}
\title{Implementing Bilinear Interpolation on Quantum Images
}


\author{Fei Yan \and Shan Zhao \and Salvador E. Venegas-Andraca}

\institute{F. Yan \Letter\at
              School of Computer Science and Technology, Changchun University of Science and Technology, China\\
              \email{yanfei@cust.edu.cn}
           \and
           S. Zhao \at
              School of Computer Science and Technology, Changchun University of Science and Technology, China\\
           \and
           S.E. Venegas-Andraca \at
              Escuela de Ingenieria y Ciencias, Tecnologico de Monterrey, Mexico\\
}

\date{Received: date / Accepted: date}

\maketitle

\begin{abstract}
In this paper, we present an interpolation scheme for FRQI images based on bilinear interpolation. To accomplish this, we formulated several quantum modules, i.e., assignment module, increment module, and quarter module, and suffused them into our proposed quantum image interpolation circuit. The concrete quantum circuits to accomplish up-scaling and down-scaling based on bilinear for FRQI images are designed and the network complexities of them are analyzed. Finally, to validate the proposed method, simulation experiments to enlarge and reduce the test images are executed, whose results are compared with the nearest neighbor interpolation for FRQI images. The up-scaled images by using proposed interpolation algorithm achieve satisfactory results, and both PSNR and SSIM values are better than those of the nearest neighbor method.

\end{abstract}

\section{Introduction}
\label{sec1}

Quantum computation is a scientific and engineering field focused on developing computers and algorithms based on quantum mechanics \cite{nielsen2000quantum,feynman1982simulating}. Quantum computing is an established field of research with solid theoretical and experimental results \cite{shor1994algorithms,grover1996fast,alanqve,gambettarmibm}  as well as  emerging fields like Quantum Machine Learning \cite{biamonteqml,schuldqml} and Quantum Image Processing (QIMP) \cite{yan2020processing}.

QIMP aims at extending classical image processing capacities on a quantum computer, being its goal to strengthen our capacity for storing, processing, and retrieving visual information from images and video either by transitioning from digital to quantum paradigms or by complementing digital imaging with quantum techniques  \cite{yan2020processing}. Quantum image representations are mathematical models used to describe how to store, process, and retrieve images in a quantum mechanical setting \cite{yan2016survey}. Similar to images, a quantum image is composed of two parts of information: color information and position information \cite{Venegas-Andraca2003storing}. Several approaches have been proposed for representing an image using a system of qubits,  among them we find the Flexible Representation of Quantum Images (FRQI) \cite{le2011flexible} and the Novel Enhanced Quantum representation (NEQR) \cite{zhang2013neqr}.  These two representations have their own advantages, e.g., FRQI uses fewer number of qubits to prepare the images and provides flexible geometric and color transformations; NEQR provides a more accurate strategy to retrieve the quantum image information and exhibits a lower complexity of similar transformations on a quantum image.

%\newpage{}
Effects on quantum images (e.g.,  image translation, rotation, and interpolation) are realized via quantum operations. Image interpolation occurs when resizing or distorting an image from one pixel grid to another. It works on both dimensions of the image and its goal is to accurately calculate the intensity of a pixel based on the values of its neighborhood (i.e., surrounding pixels) \cite{siu2012review}. Image interpolation is widely used in image enhancement \cite{Aboshosha2019Resolution}, super resolution \cite{Freeman2002example}, and other image processing techniques  \cite{luo2010reversible,upchurch2017deep,liu2019threshold}.

Interpolation algorithms for NEQR images have been studied using the nearest neighbor  \cite{jiang2015integer} and bilinear  \cite{zhou2017quantum} techniques. In these algorithms,  mathematical tools like the quantum Fourier transform are used to build quantum arithmetic modules \cite{li2018bilinear}. As for the FRQI model, interpolation algorithms have focused on the nearest neighbor interpolation approach which degrades the quality of the image due to blocking and aliasing effects  \cite{sang2016quantum}. In contrast, bilinear interpolation, where the target pixel is obtained by the linear interpolation in both horizontal and vertical directions, is capable to offer more accurate results.

In this paper, we present a quantum algorithm based on the bilinear interpolation method for FRQI images. Quantum image bilinear interpolation on FRQI images translates into two problems. Firstly, the establishment of coordinate mapping relationship between scaled images and the original images. Secondly, to calculate the corresponding intensity value of each pixel in the scaled image.

The rest of this paper is organized as follows. Section 2 presents a brief overview of the FRQI images and basic quantum operations used for bilinear interpolation. Section 3 introduces the bilinear interpolation method FRQI images. Several operations are designed to implement up-scaling and down-scaling, with their validation via quantum circuit constructions. Furthermore, we present two examples and present an analysis on circuit complexity. Section 4 shows simulation experiments to demonstrate the feasibility and effectiveness of the algorithm.

\section{Quantum preparations for images and bilinear interpolation}
\label{sec2}

\subsection{Storage and manipulation of FRQI images}
\label{subsec2.1}

The FRQI image representation model is given by
\begin{equation}
 |I\rangle  = \frac{1}{2^n}  \sum_{i=0}^{2^{2n}-1} |c_i\rangle \otimes |i\rangle,
 \label{frqidef}
\end{equation}
where $\theta = (\theta_0, \theta_1, \ldots, \theta_{2^{2n}-1} )$ is a vector of angles encoding colors, $|c_i\rangle = \cos \theta_i |0\rangle + \sin \theta_i |1\rangle$, and $|c_i\rangle$ and $|i\rangle$ respectively encode color and position in the image.

According to the polynomial preparation theorem (PPT) \cite{le2011flexible}, given an angle vector for encoding colors, $\theta=(\theta _0, \theta_1, \cdots, \theta_{2^{2n}-1})$ where $\theta_i\in [0, \frac{\pi}{2}]$, there is a unitary transformation $\mathscr{P}$ that turns an set of $2n+1$ qubits all identically initialized as $|0\rangle$ (i.e., $|0\rangle^{\otimes 2n-1}$) into Eq. (\ref{frqidef}) using a polynomial number of elementary quantum gates.
The unitary transformation  $\mathscr{P}$ is composed of the following two steps:

\begin{itemize}
	\item \textit{Step 1}: Compute  $\vert H \rangle = U_1 \vert 0 \rangle^{\otimes 2n+1} $ where  $\vert H \rangle = \frac{1}{2^n}\sum_{i= 0}^{2^{2n}-1}\vert 0\rangle\otimes\vert i\rangle$ (Eq. (\ref{eq1})).
	
	\item \textit{Step 2}: Use controlled-rotation operations to transform  $\vert H \rangle$ into the quantum image state $\vert I\rangle$.
\end{itemize}

Let $I$ and $H$ be the 2D Identity and Hadamard matrices, respectively.  We define $U_1 = I \otimes H^{\otimes{2n}}$ so that
\begin{equation}\label{eq1}\begin{aligned}
U_1 (\vert 0\rangle^{\otimes 2n+1})&= I \otimes\vert 0\rangle\otimes H^{\otimes{2n}}\otimes\vert 0\rangle^{\otimes 2n}\\
&= \frac{1}{2^n}\sum_{i= 0}^{2^{2n}-1}\vert 0\rangle\otimes\vert i\rangle\\
&= \vert H\rangle.
\end{aligned}\end{equation}

State $\vert H\rangle$ can be seen as an array whose entries are numbered out by the kets $|i\rangle$ while corresponding entry content is initialized as  $|0\rangle$. In other words, $\vert H\rangle$ can be thought of as an image whose pixels are numbered out by $|i\rangle$ and whose color value has been initialized as $|0\rangle$.

We now define operator $U_2$ (Eq. (\ref{eq2})) which will be used to assign a color value for each pixel within image $|H\rangle$,
\begin{equation}\label{eq2}
U_2 = \left(\prod_{i = 0}^{2^{2n}-1}R_i\right),
\end{equation}
where
\begin{equation}\label{eq3}
R_i = \left(I\otimes\sum_{j=0, j\neq i }^{2^{2n}-1}\vert j\rangle\langle j\vert\right) + R_y(2\theta_i)\otimes \vert i \rangle\langle i \vert,
\end{equation}\\
and
\begin{equation}\label{eq4}
R_y(2\theta _i)= \left(\begin{array}{cc}\cos\theta_i&-\sin\theta_i\\
                                     \sin\theta_i&\cos\theta_i
                                   \end{array}\right).
\end{equation}

Applying $U_2$ operator on $\vert H\rangle$ produces an FRQI image $\vert I\rangle$ as shown in Eq. (\ref{eq5}):
\begin{equation}\label{eq5}\begin{aligned}
U_2 \vert H\rangle &= \left(\prod_{i=0}^{2^{2n}-1}R_i\right)\otimes\frac{1}{2^n}\vert 0\rangle\otimes \sum_{i=0}^{2^{2n}-1}\vert i\rangle\\
&= \frac{1}{2^n}\sum_{i=0}^{2^{2n}-1}(\cos\theta_i\vert 0\rangle + \sin\theta_i\vert 1\rangle)\otimes\vert i\rangle\\
&= \frac{1}{2^n}\sum_{i=0}^{2^{2n}-1}\vert C(\theta_i)\rangle\otimes\vert i\rangle=\vert I\rangle,
\end{aligned}\end{equation}
where $\vert i\rangle$, $i=0, 1, \cdots, 2^{2n}-1$, are $2^{2n}$-dimensional computational basis states to encode the positions of an FRQI image, and $\theta =(\theta_0, \theta_1, \cdots, \theta_{2^{2n}-1})$, $\theta_i\in[0, \frac{\pi}{2}]$, is the vector of angles encoding their corresponding colors. A $2\times 2$ FRQI image with the color and position information is given in Fig. \ref{figure1}.

\addfigure{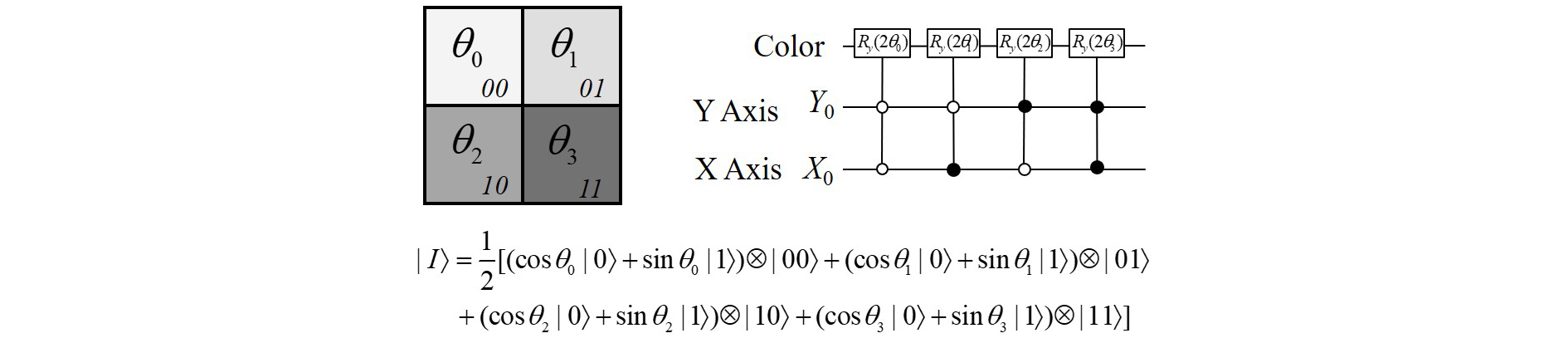}{A $2\times2$ FRQI image and its quantum circuit and state}{figure1}{width=1\textwidth}

\subsection{Quantum operations used for bilinear interpolation}
\label{subsec2.2}

In image processing, interpolation methods are applied to produce new pixels (for up-scaling) or delete redundant pixels (for down-scaling) to resize an image \cite{jiang2015quantum}. Techniques for image interpolation include nearest neighbor, bilinear, and bicubic interpolation.

Interpolation algorithms for digital images face a tradeoff: image quality vs computational complexity. For instance, let us analyze the nearest neighbour interpolation technique in which the color of target pixel in the scaled image is set as the value of the nearest pixel in the original image. Algorithms that implement nearest neighbour interpolation usually exhibit  low computational complexity but they come together with undesired effects such as mosaic and saw tooth phenomena \cite{Parker1983comparison}. Alternatively, the bicubic interpolation technique can improve the quality of scaled image but it is computationally expensive as it computes the value of each target pixel from the nearest 16 adjacent pixels \cite{han2013comparison}. Furthermore, the bilinear interpolation technique has a balanced accuracy vs computational complexity tradeoff.

In order to compute the value of each pixel in the target image, the bilinear interpolation method takes each pixel $p_i$ of the original image together with a neighborhood composed of $p_i$'s four nearest pixels located in diagonal directions  \cite{Dodgson1997Quadratic}. Suppose there is an FRQI image $\vert I\rangle$ which is considered as an original image, and the scaled image $\vert I'\rangle$ is generated from $\vert I\rangle$ by using bilinear interpolation algorithm. Assume the target pixel (to be determined) in the scaled image is $(\vert X'\rangle, \vert Y'\rangle)$, then as shown in Fig. \ref{figure2}, the color value of the target pixel is determined by four neighbor pixels $(\vert X\rangle, \vert Y\rangle)$, $(\vert X+1\rangle, \vert Y\rangle)$, $(\vert X\rangle, \vert Y+1\rangle)$, $(\vert X+1\rangle, \vert Y+1\rangle)$ in the original image. The bilinear interpolation algorithm computes the target pixel value by taking a weighted sum of the pixel values of the four nearest neighbors surrounding the target pixel.

\addfigure{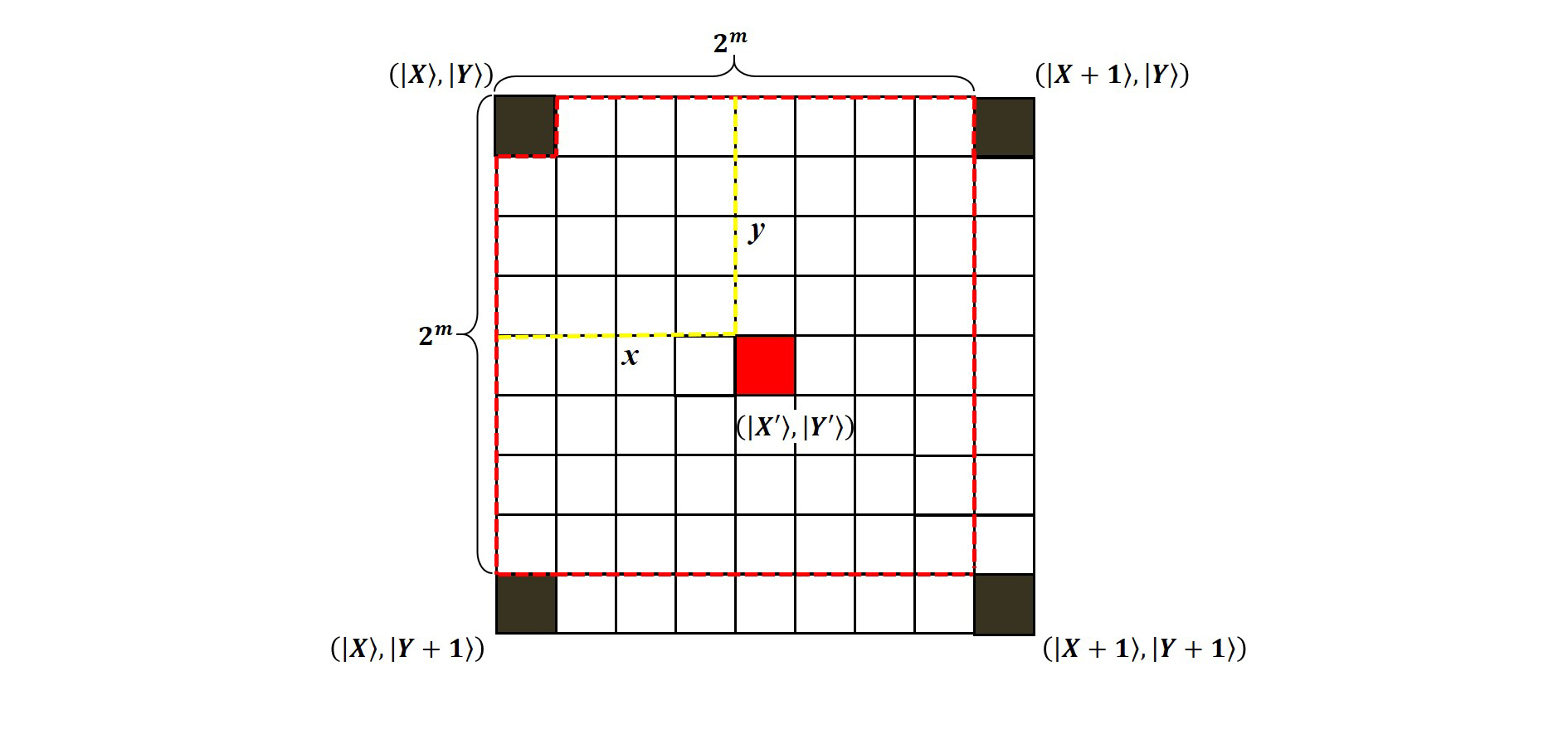}{Relationship between target pixel and its neighbor pixels}{figure2}{width=1\textwidth}

The structure of FRQI images allows applying transformation on either the full image at once or just some of its subregions \cite{le2011efficient}. Since bilinear interpolation is a local process, FRQI images are suitable for such operations.

As described earlier, since the color value of pixels in an FRQI image is encoded by a vector of angles $\theta_i=(\theta_0, \theta_1, \cdots, \theta_{2n-1})$, when a rotation gate $R_y(2\theta_k)$ is applied on the color information of the image,  the original color value is transformed according to Eqs. (\ref{eq7}) and (\ref{eq8}):
\begin{equation}\label{eq7}\begin{aligned}
R_y(2\theta_k)\vert C(\theta_i)\rangle&= \left(\begin{array}{cc}
\cos{\theta_k}&-\sin{\theta_k} \vspace{1ex} \\
\sin{\theta_k}& \cos{\theta_k} \vspace{1ex}
\end{array}\right) \left(\begin{array}{c} \cos\theta_i\\ \sin\theta_i\end{array}\right) \vspace{1ex} \\
&=\cos(\theta_i+\theta_k)\vert 0\rangle+\sin(\theta_i+\theta_k)\vert 1\rangle\\
&= \vert C(\theta_i+\theta_k)\rangle,
\end{aligned}\end{equation}
\begin{equation}\label{eq8}\begin{aligned}
R_y(-2\theta_k)\vert C(\theta_i)\rangle&= \left(\begin{array}{cc}
\cos{\theta_k}& \sin{\theta_k} \vspace{1ex} \\
-\sin{\theta_k}& \cos{\theta_k} \vspace{1ex}
\end{array}\right) \left(\begin{array}{c} \cos\theta_i\\ \sin\theta_i\end{array}\right) \vspace{1ex} \\
&=\cos(\theta_i-\theta_k)\vert 0\rangle+\sin(\theta_i-\theta_k)\vert 1\rangle\\
&= \vert C(\theta_i-\theta_k)\rangle.
\end{aligned}\end{equation}

From the equations above, it can be concluded that the effect of applying $R_y(2\theta_k)$ on the color wire is to change the color which is represented by $\theta_i$ to $\theta_i+\theta_k$, and the effect of  $R_y(-2\theta_k)$ is to change $\theta_i$ to $\theta_i-\theta_k$, i.e., $R_y(2\theta_k)$ provides a way to transform the color information of the pixels in an FRQI image. In order to constrain the color transformation within a subarea of a quantum image, a transformation method with additional control conditions on the position wires are required  \cite{le2011efficient}.

\section{Quantum realization of bilinear interpolation}
\label{sec3}

In this section, we discuss how to use bilinear interpolation to enlarge and reduce an FRQI image. Additionally, the concrete quantum circuits of bilinear interpolation are designed to realize corresponding operations (up-scaling and down-scaling) and the complexities of them are analyzed. Finally, the proposed method is illustrated by two simple examples.

\subsection{Analysis of quantum bilinear interpolation method}
\label{subsec3.1}

Let the size of the original image $\vert I\rangle$ be $ 2^n\times 2^n$, and the size of the scaled image $\vert I'\rangle$ based on bilinear interpolation method is $ 2^{n+m}\times2^{n+m}$, i.e., the scaling ratio is $r_y \times r_x= 2^m\times2^m$. To realize bilinear interpolation on FRQI images, we must address the following two key problems (we will also discuss how to use quantum circuits to compute arithmetic operations.) To calculate the value of the target pixel requires first locating the four grid points surrounding it.

%\newpage{}
\textbf{Problem 1:} Building (\textit{x}, \textit{y}) coordinate mappings from original FRQI images to corresponding scaled FRQI images.

Assume the coordinate of the target pixel is ($\vert X'\rangle, \vert Y'\rangle$), the horizontal axis information can be encoded by $\vert X'\rangle = \vert X'_{n+m-1}X'_{n+m-2}\cdots X'_0\rangle$, and $\vert Y'\rangle = \vert Y'_{n+m-1}Y'_{n+m-2}\cdots Y'_0\rangle$ encodes the vertical coordinate information. As shown in Fig. \ref{figure2}, $x$ and $y$ that range from 0 to $2^m-1$ denote the distance from each target pixel to the neighbor pixel $(\vert X\rangle, \vert Y\rangle)$ in the X-axis and the Y-axis, respectively. The width from $(\vert X\rangle, \vert Y\rangle)$ to $(\vert X+1\rangle, \vert Y\rangle)$ and the height from $(\vert X\rangle, \vert Y\rangle)$ to $(\vert X\rangle, \vert Y+1\rangle)$ are both $2^m$. Thus, the corresponding relation between the neighbor pixel and target pixel can be described as follows:
\begin{equation}\label{eq9}
 \vert X'\rangle = \vert\frac{ X}{2^m}+x\rangle, \vert Y'\rangle=\vert\frac{Y}{2^m}+y\rangle.
\end{equation}

In the up-scaling process, each pixel in original image are mapped into an $2^m\times 2^m$ sub area, any pixel in the sub area has same neighbor pixels in the original image. Equation (\ref{eq9}) can be realized by using a sequence of CNOT gates in the quantum circuit (we call it position-assignment (PA) module), i.e., copy the first $n$ qubits of the target pixel $\vert X'\rangle=\vert X'_{n+m-1}X'_{n+m-2}\cdots X'_0\rangle$ into $n$ auxiliary qubits $\vert 0\rangle^{\otimes n}$ by using $n$ CNOT gates as shown in Fig. \ref{figure3}. As such, through the position information of the target pixel ($\vert X'\rangle, \vert Y'\rangle$), its neighbor pixel ($\vert X\rangle, \vert Y\rangle$) can be found in the original image.

\addfigure{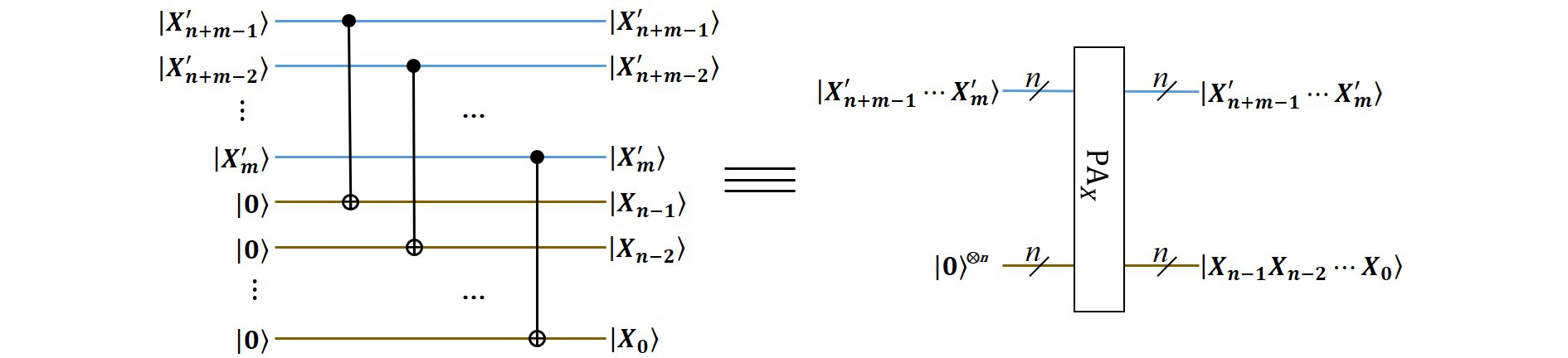}{Quantum circuit of the PA module} {figure3}{width=1\textwidth}

Now, we proceed to find the coordinate information of the other pixels $(\vert X+1\rangle, \vert Y\rangle)$, $(\vert X\rangle, \vert Y+1\rangle)$, and $(\vert X+1\rangle, \vert Y+1\rangle)$. To facilitate this, in addition to the aforementioned PA modules, some additional operations, e.g., the increment module $\Omega$, are required as shown in Fig. \ref{figure4}. In Fig. \ref{figure4}, when $X_{n-1}\times X_{n-2}\times\cdots\times X_0 \neq 1$, the equation $\Omega \vert X\rangle = \vert X+1\rangle$ can be realized. Otherwise, when all the input qubits in Fig. \ref{figure4} are ``1'', the current pixel is at the edge of the original image (e.g., the pixels outside the red rectangle in Fig. \ref{figure2}). Since the position $\vert X+1\rangle$ can't be found in the original image, we define $\Omega \vert X\rangle = \vert X\rangle$ to avoid the overflow of the image size. Therefore, when the pixel coordinate $(\vert X\rangle, \vert Y\rangle)$ is known, all the other neighbor pixels of the target pixel in the scaled image $(\vert X+1\rangle, \vert Y\rangle)$, $(\vert X\rangle, \vert Y+1\rangle)$, and $(\vert X+1\rangle, \vert Y+1\rangle)$ can be obtained successfully according to the $\Omega$ modules.

\addfigure{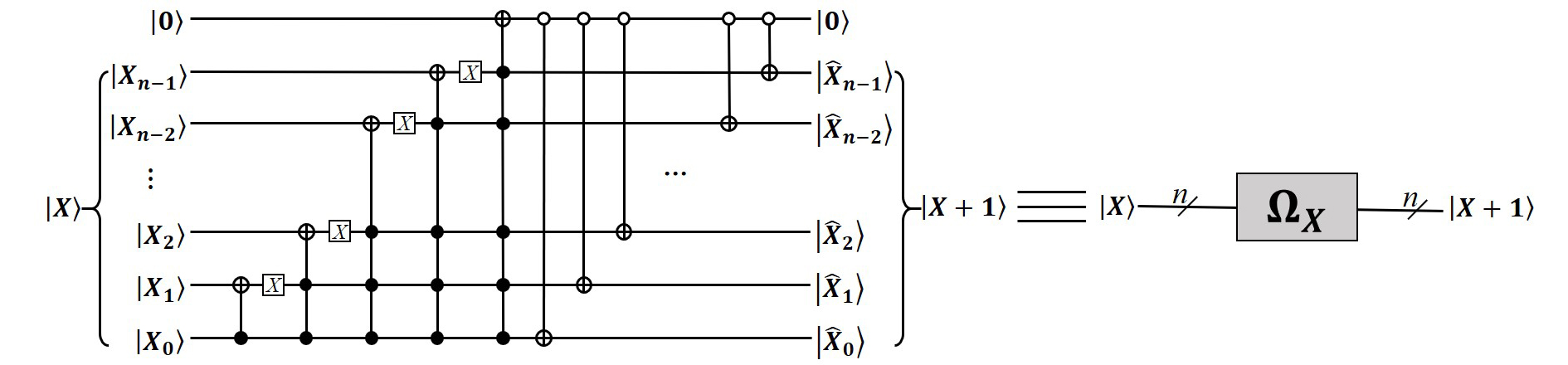}{Quantum circuit of the $\Omega$ module} {figure4}{width=1\textwidth}

\textbf{Problem 2:} How to calculate the gray value of the target pixel?

Bilinear interpolation considers the closest $2 \times 2$ neighborhood of known pixel values surrounding the unknown pixel's computed location. The intensity value of the interpolated pixel is equal to the weighted sum of the pixels that compose the $2 \times 2$ neighborhood already mentioned.

So, the gray value of the target pixel can be obtained according to Eq. (\ref{eq10}):
\begin{equation}
\label{eq10}
\begin{aligned}
\vert f(X', Y')\rangle&=\omega_1 \vert f(X, Y)\rangle +\omega_2 \vert f(X+1, Y)\rangle\\
 &+\omega_3 \vert f(X, Y+1)\rangle+\omega_4 \vert f(X+1, Y+1)\rangle.
\end{aligned}
\end{equation}

Now, the weights required for the computation of Eq. (\ref{eq10}) are determined by their distances from the target pixel, which is defined according to Eq. (\ref{eq11}):
\begin{equation}
\label{eq11}
\begin{aligned}
\omega_1 &= (2^m-x)(2^m-y)/{2^{2m}},\\
\omega_2 &= (2^m-x)/{2^{m}},\\
\omega_3 &= (2^m-y)/{2^{m}},\\
\omega_4 &= xy/{2^{2m}}.
\end{aligned}
\end{equation}

Repeating this process on all pixels of the original image yields the full scaled image. We now discuss how the pixel values of scaled images are computed.

The color-assignment (CA) module circuit (Fig. \ref{figure5}) is designed by using a series of controlled-rotation gates, which can be used to assign the gray value of the target pixel based on the gray value of the known pixels. If the coordinates of the neighbor pixel are $(\vert X\rangle, \vert Y\rangle)$ then the output $\vert f(X, Y)\rangle$  is computed by the CA-1 module presented in Fig. \ref{figure5}.

\addfigure{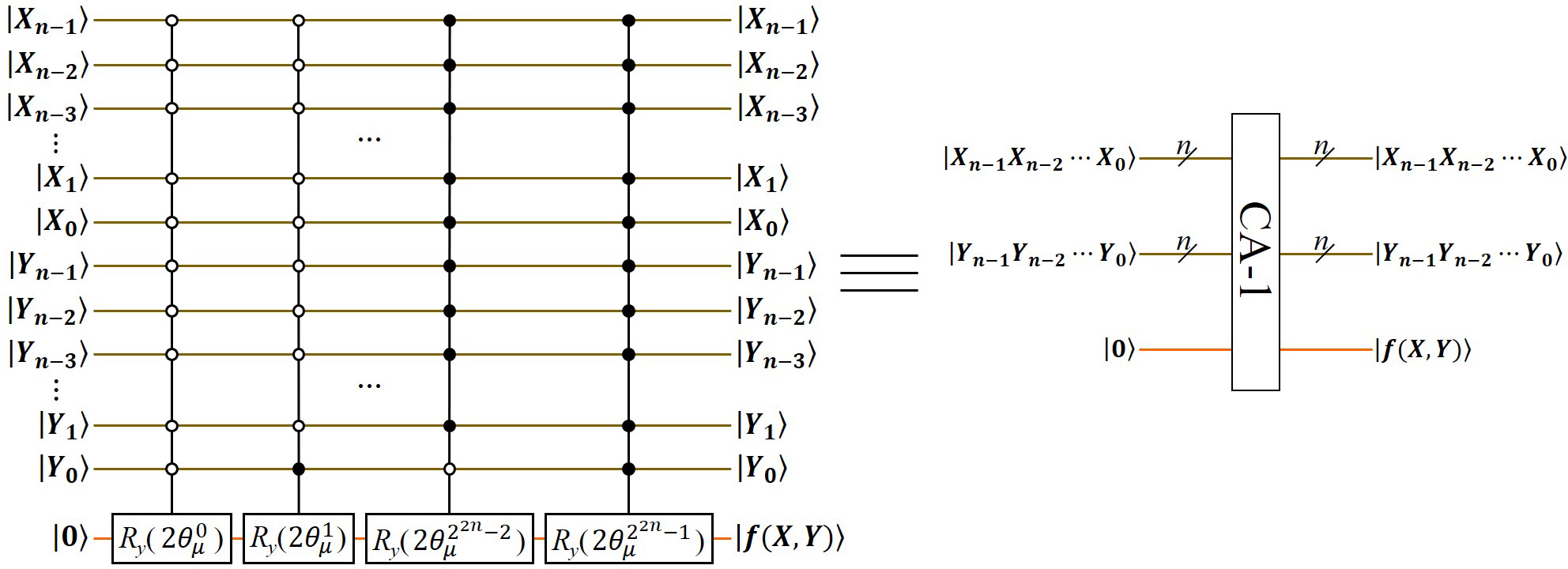}{Quantum circuit of the CA module}{figure5}{width=1\textwidth}

As mentioned above, each neighbor pixel requires a set of weight coefficients which are employed to compute the final gray value of the target pixel. We present the weight-assignment (WA) module in Fig. \ref{figure6}.  The angle parameter operator of the rotation operation denoted as ($\theta_{\hat{\mu}}$, $\theta_{\hat{\nu}}$, $\theta_{\hat{\tau}}$, $\theta_{\hat{\epsilon}}$) are set by
\begin{equation}
\label{eq12}
\begin{aligned}
\omega_1\theta_{\mu}&= \theta_\mu +\theta_{\hat{\mu}}&\Rightarrow \theta_{\hat{\mu}}&=(\omega_1-1)\theta_{\mu},\\
\omega_2\theta_{\nu}&= \theta_\nu +\theta_{\hat{\nu}}&\Rightarrow \theta_{\hat{\nu}}&=(\omega_2-1)\theta_{\nu},\\
\omega_3\theta_{\tau}&= \theta_\tau +\theta_{\hat{\tau}}&\Rightarrow \theta_{\hat{\tau}}&=(\omega_3-1)\theta_{\tau},\\
\omega_4\theta_{\epsilon}&= \theta_\epsilon +\theta_{\hat{\epsilon}}&\Rightarrow \theta_{\hat{\epsilon}}&=(\omega_4-1)\theta_{\epsilon},
\end{aligned}
\end{equation}
where ($\theta_\mu$, $\theta_\nu$, $\theta_\tau$, $\theta_\epsilon$)  are the vectors of angles encoding color information of neighbor pixels $(\vert X\rangle, \vert Y\rangle)$, $(\vert X+1\rangle, \vert Y\rangle)$, $(\vert X\rangle, \vert Y+1\rangle)$, $(\vert X+1\rangle, \vert Y+1\rangle)$. Thus for each neighbor pixel, it is always possible to find a rotation angle to transform its gray value information to a weighted value. Then, the gray value of the target pixel $\vert f(X', Y')\rangle$ at each pixel in the scaled image can be obtained.

\addfigure{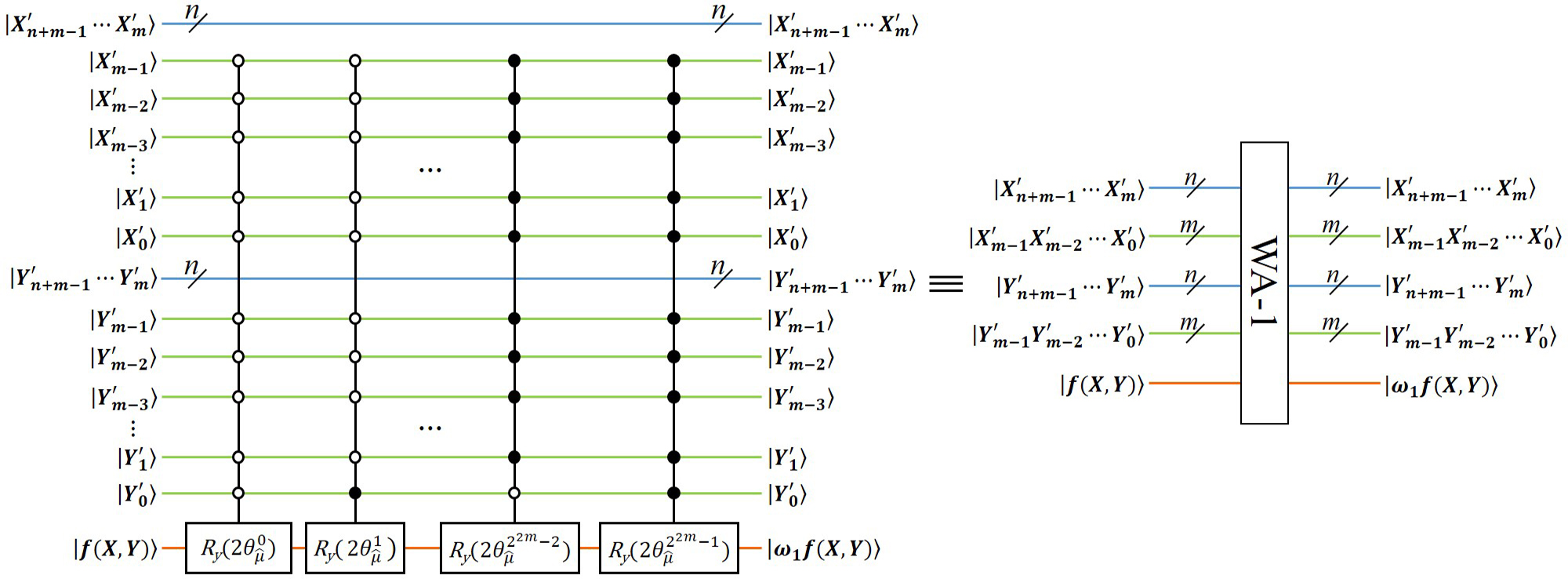}{Quantum circuit of the WA module}{figure6}{width=1\textwidth}

\subsection{Quantum circuit realization of up-scaling operation}
\label{subsec3.2}

Based on the circuits presented above, we introduce in this section a quantum circuit to realize up-scaling images by using the bilinear interpolation (Fig. \ref{figure7}). Our up-scaling procedure  is described in the following  three steps.

\begin{itemize}
\item \textit{Step 1}: We assign the values of the first $2n$ qubits of the target pixel's coordinate $(\vert X'\rangle, \vert Y'\rangle)$ to the $2n$ initialized quantum states $\vert 0\rangle^{\otimes2n}$ by using the PA modules (as discussed in Fig. \ref{figure3}), so as to find the coordinate information of the corresponding neighbor pixel $(\vert X\rangle, \vert Y\rangle)$ in the original image.

\item \textit{Step 2}: Using the PA and $\Omega$ modules presented in Fig. \ref{figure4}, we compute the positions of the three neighbor pixels of $(\vert X\rangle, \vert Y\rangle)$.

\item \textit{Step 3}: We run the CA-1 module presented in Fig. \ref{figure5}  to compute the gray value of pixel($\vert X\rangle, \vert Y\rangle$). Then, the WA-1 module (Fig. \ref{figure6}) is applied to calculate its weighted value. By using the swap gate to retain the earlier result from the WA-1 module, the CA-2 and WA-2 operations compute the sum of weighted gray value of ($\vert X\rangle,  \vert Y\rangle$) and ($\vert X+1\rangle, \vert Y\rangle$). By repeating this, the final weighted sum of the four neighbor pixels, i.e., the gray value of($\vert X'\rangle, \vert Y'\rangle$), is computed.
\end{itemize}

\addfigure{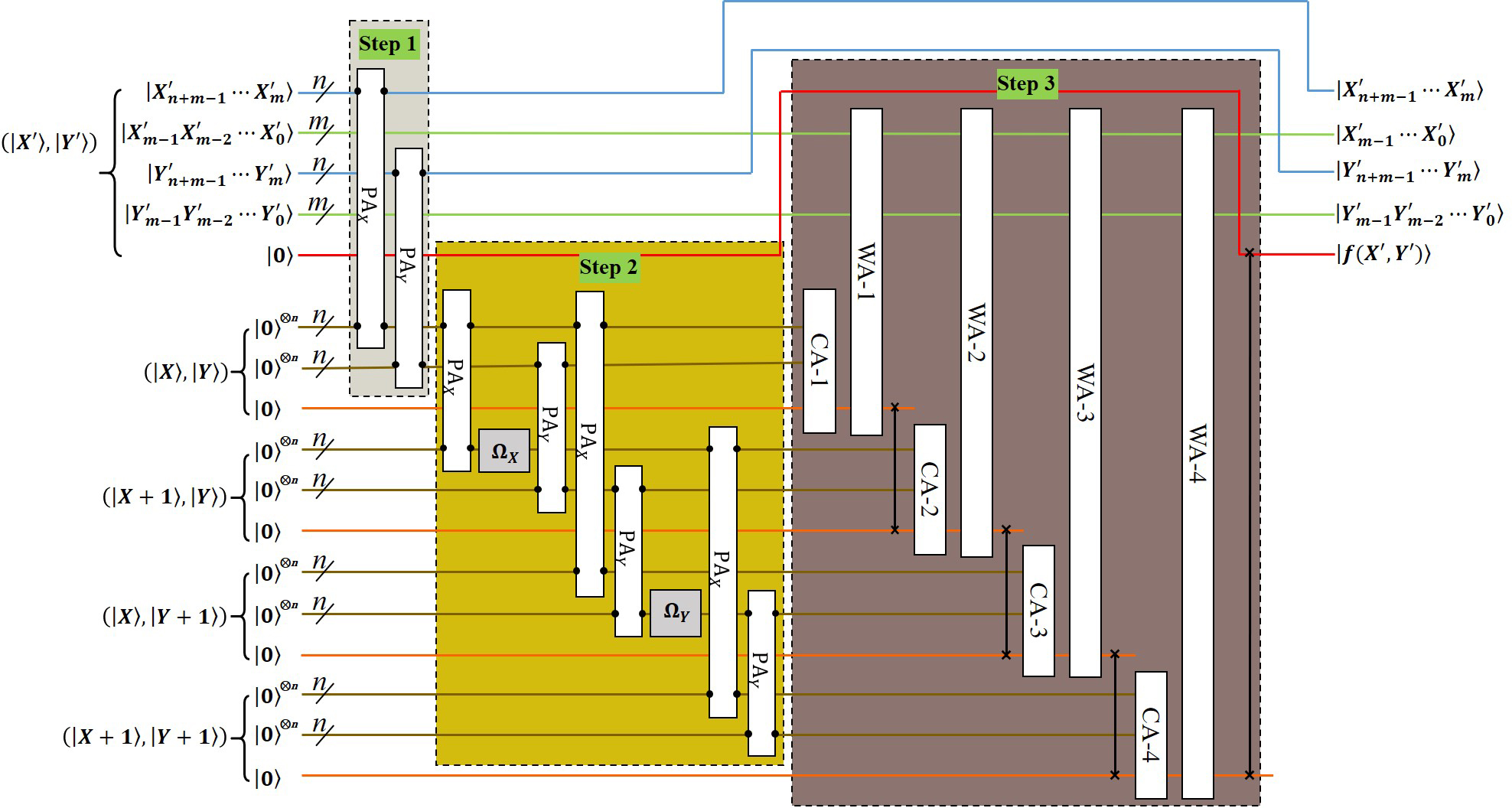}{The up-scaling circuit based on bilinear interpolation for FRQI images}{figure7}{width=1\textwidth}

From the steps described above, the up-scaled image $\vert I'\rangle$ can be fully obtained from the original image $\vert I\rangle$. In the following lines, we present an example of up-scaling starting from a  $2^2\times2^2$ gray scale image.

The scaling ratio is $2^1\times 2^1$ and the up-scaling process is shown in Fig. \ref{figure8}. For example, the pixel $(\vert X'\rangle, \vert Y'\rangle)$=$(\vert 010\rangle, \vert 101\rangle)$) marked in a red box on the right of Fig. \ref{figure8}, whose gray value is calculated from the four neighbor pixels $(\vert 01\rangle, \vert 10\rangle)$, $(\vert 10\rangle, \vert 10\rangle)$, $(\vert 01\rangle, \vert 11\rangle)$, and $(\vert 10\rangle, \vert 11\rangle)$ marked in the yellow box on the left image of Fig. \ref{figure8}.

\addfigure{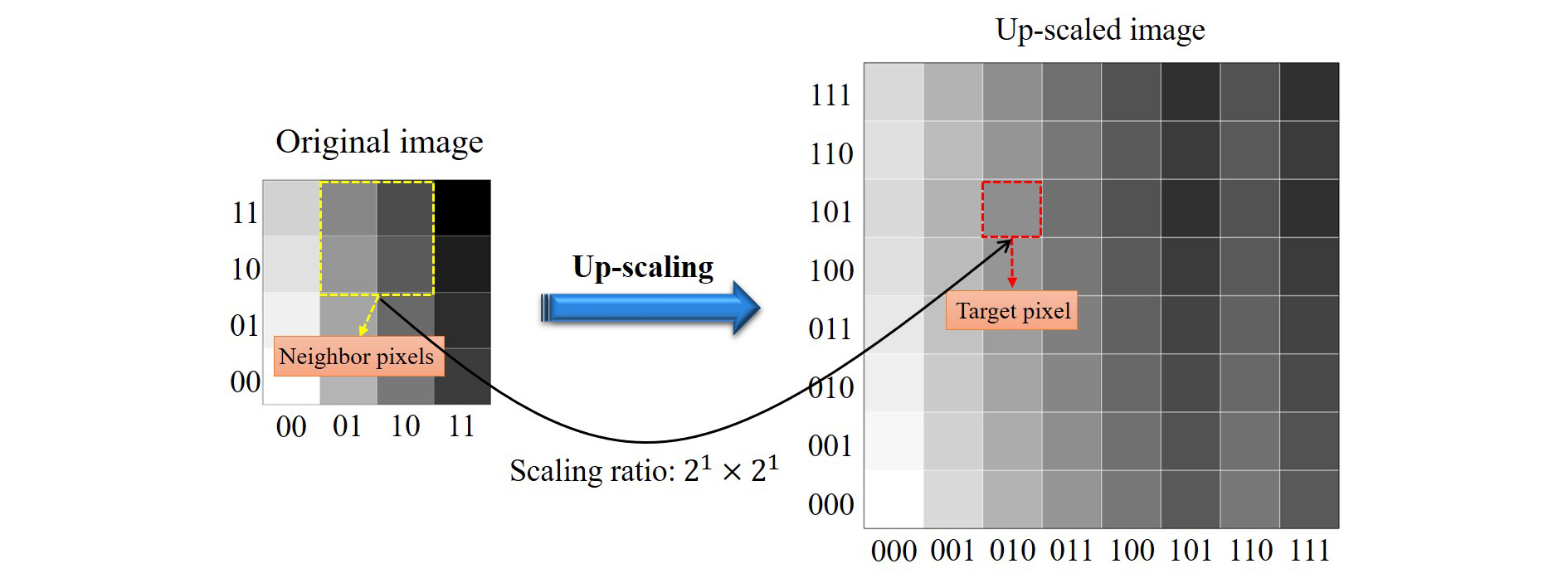}{A up-scaling example based on bilinear interpolation for FRQI images}{figure8}{width=1\textwidth}

Now, the complexity of the proposed up-scaling circuit for FRQI images will be discussed. Classical complexity of  image processing is usually computed in terms of algorithm running time on classical computer. Since quantum gates play a central role in many quantum computational networks, according to the complexity theory on quantum computation, the complexity of quantum algorithms depends very much on the number of the elementary gates \cite{Vlatko1996quantum}. Thus, the quantum circuit of the operations used in the up-scaling mechanism proposed in this study, will be decomposed by employing the strategy for circuit decomposition presented in \cite{barenco1995elementary}. A visual help for the following computation can be found in Fig. \ref{figure7}, in which we present the  PA  module (firstly introduced in Fig. \ref{figure3}), $\Omega$ module (also presented in Fig. \ref{figure4}), CA modules (also presented in Fig. \ref{figure5}), WA modules (firstly introduced in Fig. \ref{figure6}), and a series of Swap gates, when we consider the CNOT gate as the basic computing unit, the complexities can be analyzed as follows.

\begin{itemize}
\item\textbf {PA module:}\\
The PA module consists of $n$ CNOT gates. There are eight PA operations in Step 1 and Step 2 thus the complexity of the eight PA operations is $8n$.

\item \textbf{$\Omega$ module:}\\
The $\Omega$ module in Step 2 can be run by $n-1$ NOT, $n+1$ CNOT, 1 Toffoli, and $\underline{n-2}$ $k$-controlled NOT gates, where $(k = 3, 4, \cdots, n)$. The decomposition of the $k$-controlled NOT gate into simpler circuit networks is illustrated in Fig. \ref{figure9}.

\addfigure{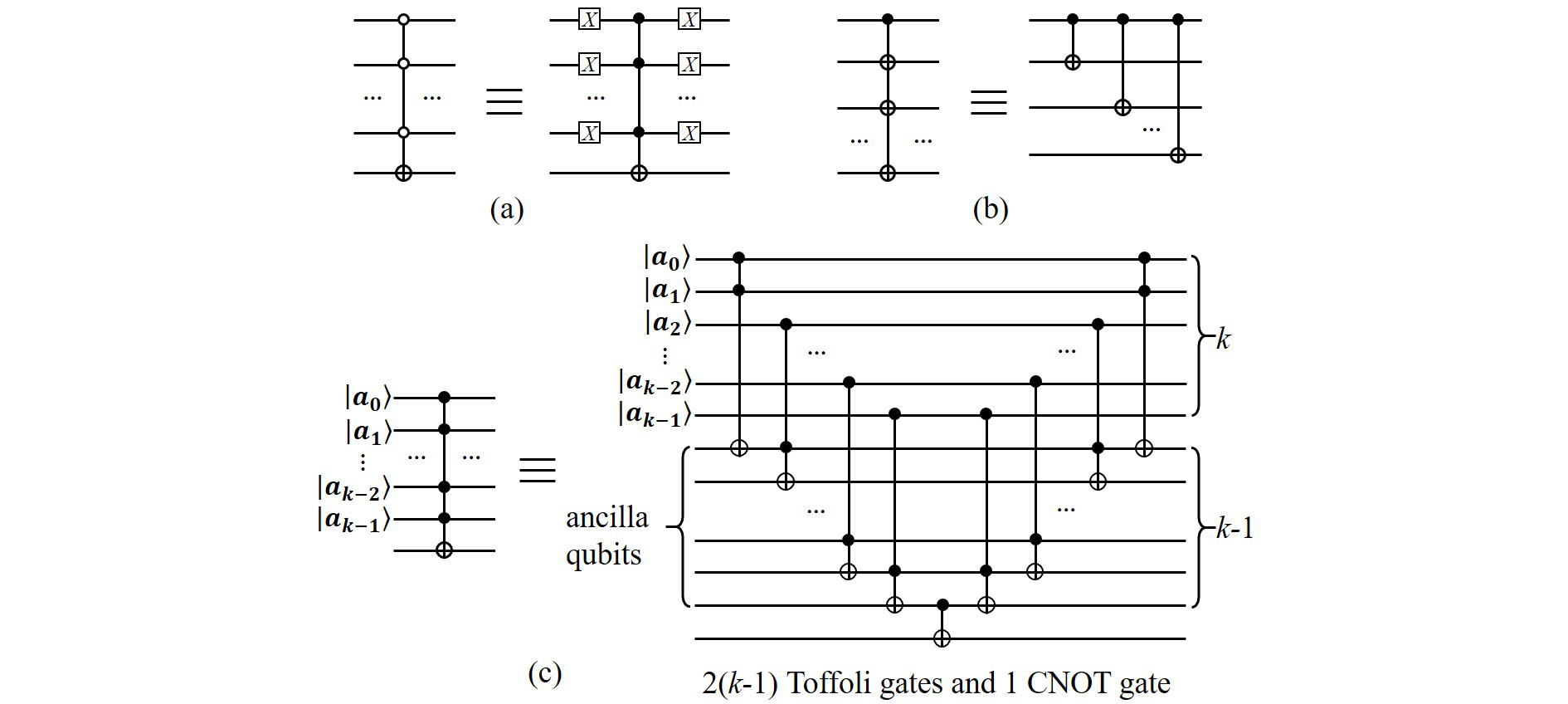}{Decomposition of $k$-controlled NOT gate into network of basic gates}{figure9}{width=1\textwidth}

As shown in Fig. \ref{figure9}, a $k$-controlled NOT gate can be decomposed into $2(k-1)$ Toffoli gates and 1 CNOT gate. Additionally, one Toffoli gate can be further approximately simulated by six CNOT gates \cite{barenco1995elementary}. Hence, the complexity of a $k$-controlled NOT gate is $12k-11$. For all $k$-controlled NOT gates, the complexity is given by
\begin{equation}
\label{eq13}
\sum_{k=3}^{n}(12k-11)= 6n^2-5n-14.
\end{equation}

Therefore, the complexity of the $\Omega$ module can be calculated as
\begin{equation}
\label{eq14}
\begin{aligned}
O(\Omega) &=n+1+6+6n^2-5n-14\\
&=6n^2-4n-7.
\end{aligned}
\end{equation}

\item \textbf{CA module:}\\
A CA module in Fig. \ref{figure5} contains $2^{2n}$ $2n$-controlled rotation gates. According to \cite{barenco1995elementary}, a $2n$-controlled rotation gates can be broken down into two controlled-rotation gates and two $(2n-1)$-controlled NOT gates. So, the complexity of the CA module is
\begin{equation}
\label{eq15}
\begin{aligned}
O(\rm CA) &= 2\times\Big(2^{2n}+12\big[(2n-1)-11\big]\Big)\times 2^{2n}\\
&\approx 2^{4n}.
\end{aligned}
\end{equation}

\addfigure{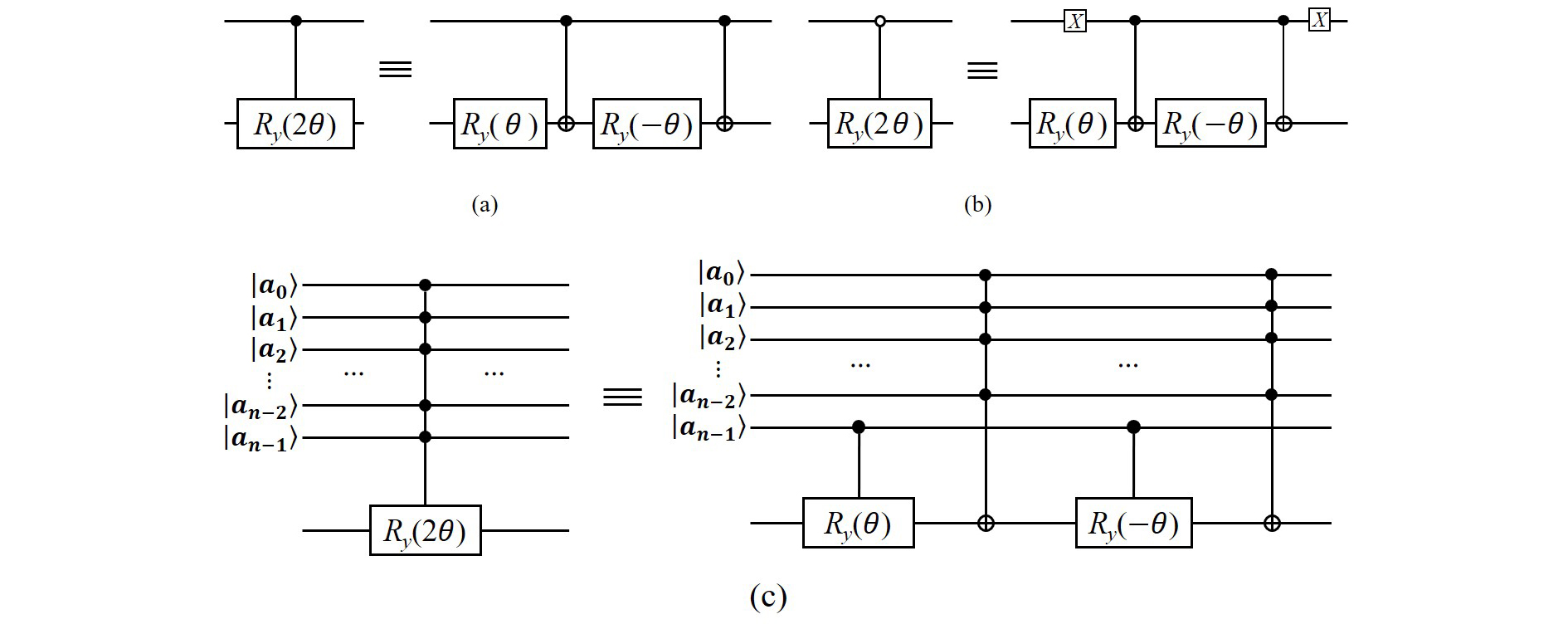}{Decomposition of controlled-rotation gate into network of basic gates}{figure10}{width=1\textwidth}

\item \textbf{WA module:}\\
The WA module is similar to the CA module, as shown in Fig. \ref{figure6}, the number of the control qubits is $2m$. Then, the complexity of a WA module is
\begin{equation}
\label{eq16}
\begin{aligned}
O(\rm WA) &= 2\times\Big(2^{2m}+12\big[(2m-1)-11\big]\Big)\times 2^{2m} \\
&\approx 2^{4m}.
\end{aligned}
\end{equation}

\item \textbf{ Swap gate:}\\
One Swap gate is equivalent to three CNOT gates as shown in Fig. \ref{figure11}.

\addfigure{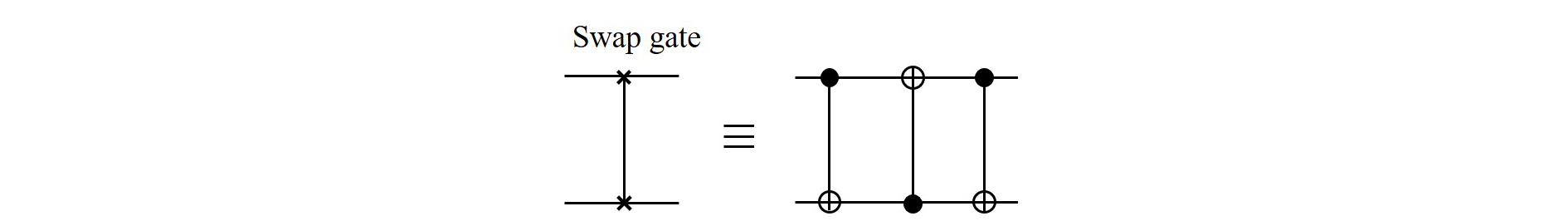}{Decomposition of a Swap gate into network of basic gates}{figure11}{width=1 \textwidth}
\end{itemize}

So, when an FRQI image with a size of $2^n \times 2^n$ is enlarged to an image with a size of $2^{n+ m} \times 2^{n+ m}$ by bilinear interpolation, the complexity of the whole up-scaling circuit as presented in Fig. \ref{figure7} can be calculated as
\begin{equation}
\label{eq17}
\begin{aligned}
O(\rm {up\mbox{-}scaling}) &=8\rm PA+2\Omega+4CA+4WA+4Swap \\\
&=8n+2\times(6n^2-4n-7)+4\times 2^{4n}+4\times 2^{4m}+4\times3\\
&=12n^2-2+2^{4n+2}+2^{4m+2}\\
&\approx 2^{4n}+2^{4m}.
\end{aligned}
\end{equation}

It is worth noting that the size of an image, $2^n$, {\it is not meant to asymptotically grow exponentially large as $n$ is a fixed number, not a function}. Modern digital images produced by professional cameras are of the order of megapixels. For instance, the $\alpha 9^{\tiny{\textregistered}}$ Sony$^{\tiny{\textregistered}}$ camera has a 24.2 megapixel resolution, that is $2.42 \times 10^7$ pixels per digital photograph. Thus, $m^2 = 2.42 \times 10^7 \Rightarrow m \approx 4920$. Since $2^{12} = 4096$ and $ 2^{13} = 8192$, we may set $r= 13$ as an upper bound for modern digital camera technology.

\subsection{Quantum circuit realization of down-scaling operation}
\label{subsec3.3}

Assume the size of original image $\vert I\rangle$ is $2^{n+m}\times 2^{n+m}$, we aim to reduce it to $\vert I'\rangle$ which is a $2^n\times2^n$-sized image. Hence, the scaling ratio is $r_y \times r_x = 2^{-m}\times2^{-m}$. The gray value of the pixel in the down-scaled image can be calculated by an average of the gray values of four selected pixels in the original image. We aim to use the $(\vert X'\rangle, \vert Y'\rangle)$ in the down-scaled image to find the four pixels in the original image. Considering the accuracy and the complexity of the circuit design, we propose a new method to define the positions of the four pixels as $P_1=\big(\vert X'\cdot 2^m\rangle, \vert Y'\cdot 2^m\rangle\big)$, $P_2=\big(\vert (X'+1)\cdot 2^m\rangle, \vert Y'\cdot 2^m\rangle\big)$, $P_3=\big(\vert X'\cdot 2^m\rangle, \vert (Y'+1)\cdot 2^m\rangle\big)$, and $P_4=\big(\vert (X'+1)\cdot 2^m\rangle, \vert (Y'+1)\cdot 2^m\rangle\big)$. The gray value of the pixel in the down-scaled image based on bilinear interpolation is defined as
\begin{equation}
\label{eq18}
\begin{aligned}
\vert f(X', Y')\rangle &=\frac{1}{4}\Big(\vert f(P_1)\rangle+\vert f(P_2)\rangle+\vert f(P_3)\rangle+\vert f(P_4)\rangle\Big),
\end{aligned}
\end{equation}
where the $\vert f(X', Y')\rangle$ denotes the gray value of the target pixel in the down-scaled image, and the $\vert f(P_1)\rangle$, $\vert f(P_2)\rangle$, $\vert f(P_3)\rangle$, and $\vert f(P_4)\rangle$ denote the gray values of the four selected pixels in the original image. The quantum circuit of bilinear interpolation is designed, which is presented in Fig. \ref{figure12}.

\addfigure{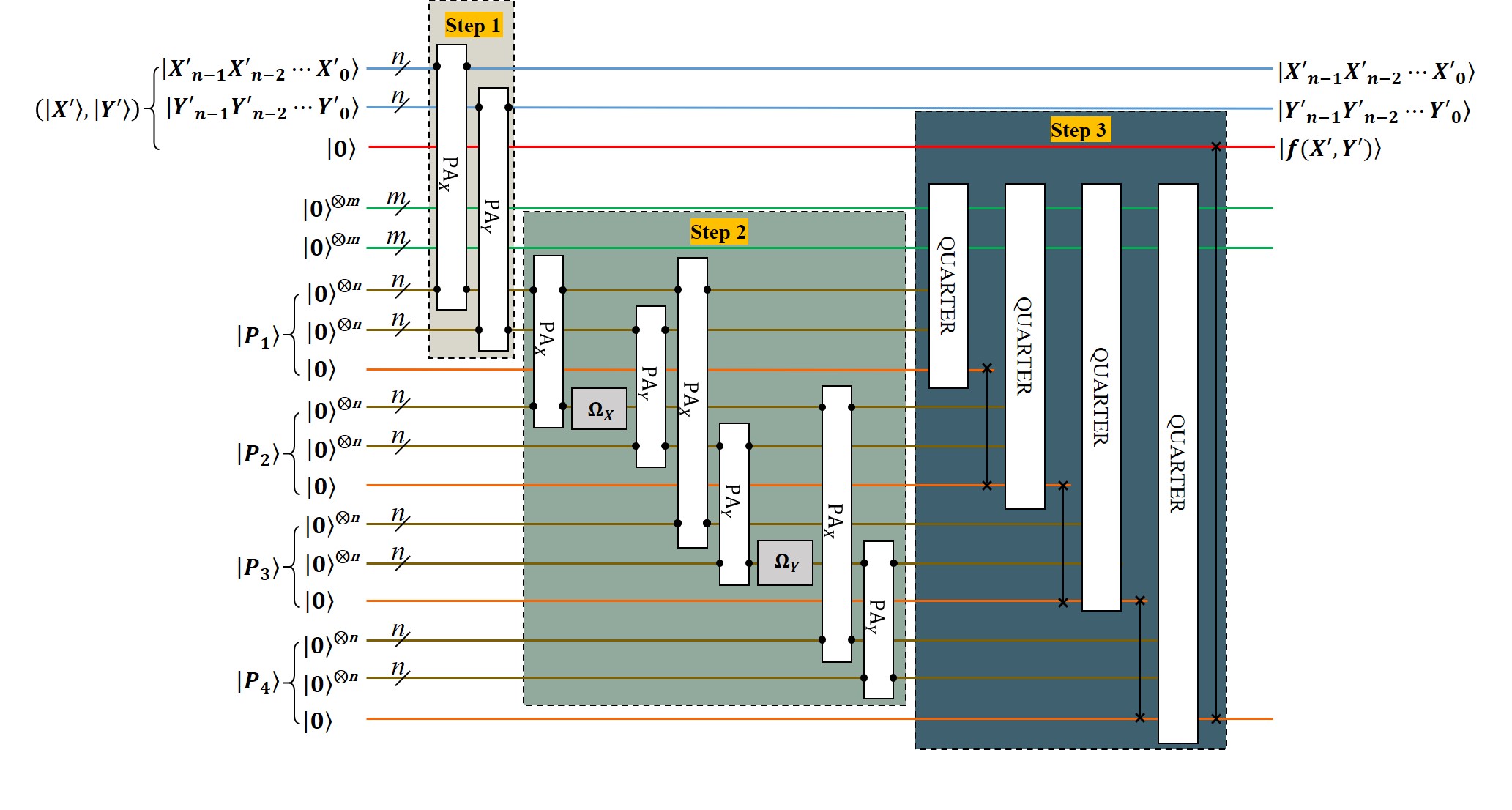}{The down-scaling circuit based on bilinear interpolation for FRQI images}{figure12}{width=1.0\textwidth}

\newpage{}

Following this, the down-scaling procedure can be accomplished by the following three steps.

\begin{itemize}
\item \textit{Step 1}: By using the PA modules, we can assign the coordinate information of the target pixel $(\vert X'\rangle, \vert Y'\rangle)$ in the down-scaled image $\vert I'\rangle$ to initial quantum states $\vert 0\rangle^{\otimes{2n}}$, so as to find the  coordinates of pixel $\vert P_1\rangle$ in the original image $\vert I\rangle$. It is noteworthy that $\vert I\rangle$ include $2n+2m$ qubits for its coordinate information. To reduce the complexity, we set its latter $2m$ qubits as $\vert 0\rangle^{\otimes{2m}}$.

\item \textit{Step 2}: After the additional PA operations, the $\Omega_X$ module is applied on the first $n$ qubits (rather than the whole $n+m$ qubits) of the $X$ coordinate of the pixel $\vert P_2\rangle$, so the coordinate information of the pixel $\vert P_2\rangle$ is obtained. Since the latter $2m$ qubits of the four selected pixels are all set as $\vert 0\rangle^{\otimes{2m}}$, for brevity, these qubits are drawn together with green wires in the circuit for the calculation. In a similar way, the coordinate information of $\vert P_3\rangle$ and $\vert P_4\rangle$ could be obtained.

\item \textit{Step 3}: We apply the QUARTER module (Fig. \ref{figure13}) on the color wires of the pixel $\vert P_1\rangle$, then with the Swap gates, one fourth of its gray value will be obtained and further accumulated to the next pixel. By repeating this, finally, the gray value of the target pixel $(\vert X'\rangle, \vert Y'\rangle)$ is obtained in the down-scaled image.
\end{itemize}

\addfigure{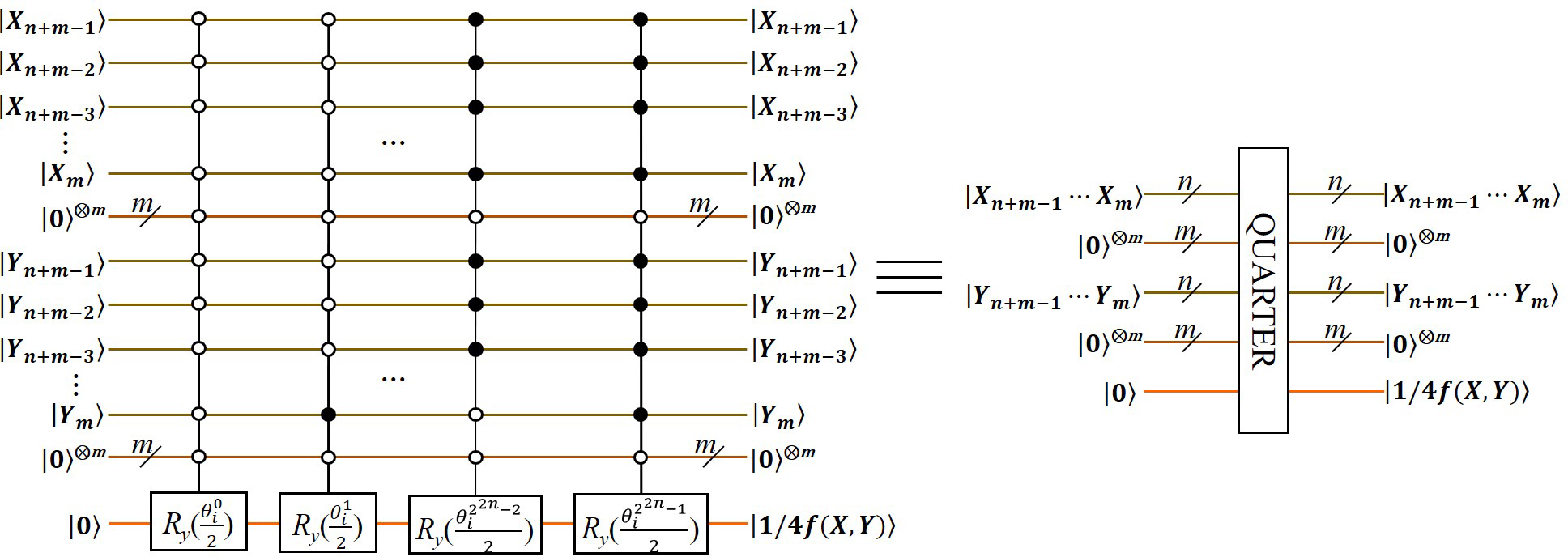}{Quantum circuit of the QUARTER module}{figure13}{width=1\textwidth}

We now present an example of our down-scaling process. A $2^3\times 2^3$ FRQI image is used for this purpose (Fig. \ref{figure14}) and the scaling ratio is set as $2^{-1}\times 2^{-1}$. The pixel in the down-scaled image is $(\vert X'\rangle, \vert Y'\rangle)$=$(\vert01\rangle, \vert 10\rangle)$  (marked in a red box in the image on the right of Fig. \ref{figure14}) and whose gray value is calculated from the four selected pixels in original image $(\vert 010\rangle, \vert 100\rangle)$, $(\vert 100\rangle, \vert 100\rangle)$, $(\vert 010\rangle, \vert 110\rangle)$, and $(\vert 100\rangle, \vert 110\rangle)$ marked in the yellow box on the left of the figure.

\addfigure{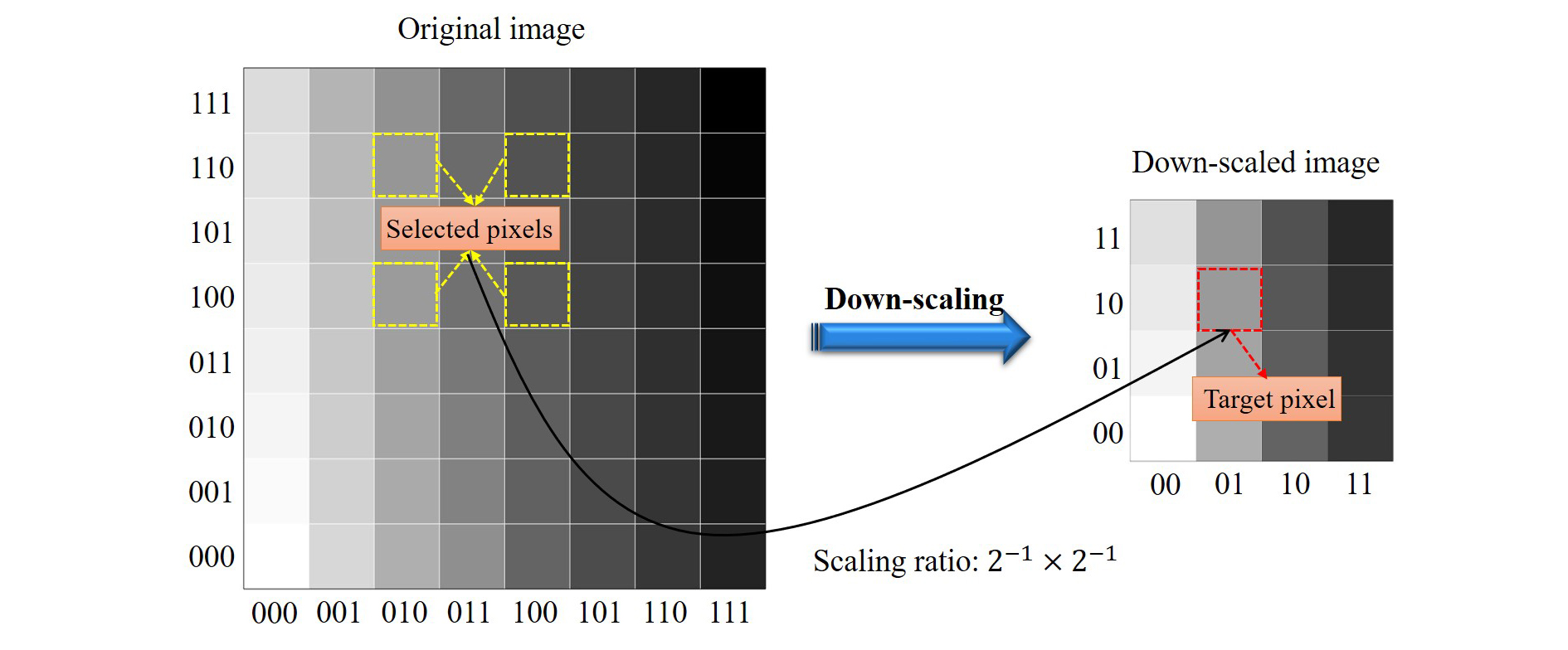}{A down-scaling example based on bilinear interpolation for FRQI images}{figure14}{width=1\textwidth}

We now analyze the complexity of our down-scaling circuit. As Fig. \ref{figure12} shows, the down-scaling circuit contains eight $n$-controlled NOT gates, two $\Omega$ modules acting on $n$ qubits, four QUARTER modules and four Swap gates. The complexity of the PA and $\Omega$ modules, and Swap gates have been analyzed in Section \ref{subsec3.3}. Since the QUARTER module is constructed by $\underline{2^{2n}}$ $2(m+n)$-controlled rotation gates, the complexity of QUARTER module is
\begin{equation}
\label{eq19}
\begin{aligned}
O(\rm QUARTER) &=2^{2n}\times2\times\Big(2+12\big[2(m+n)-11\big]\Big)\\
&=2^{2n}\times\big[48(m+n)-18\big]\\
&\approx 2^{2n}.
\end{aligned}
\end{equation}

Furthermore, the complexity of the full down-scaling circuit is
\begin{equation}
\label{eq20}
\begin{aligned}
O(\rm down\mbox{-}scaling) &=8\rm PA+2\Omega+4QUARTER+4Swap\\
&=8n+2\times(6n^2-3n-8)+ 4\times2^{2n}+4\times 3\\
&\approx 2^{2n}.
\end{aligned}
\end{equation}

\section{Simulation experiment of quantum bilinear operations of FRQI images}
\label{sec4}

\begin{figure}[htbp]
    \centering
    \subfigure[$2^9\times2^9$]{
       \includegraphics[width=0.9in]{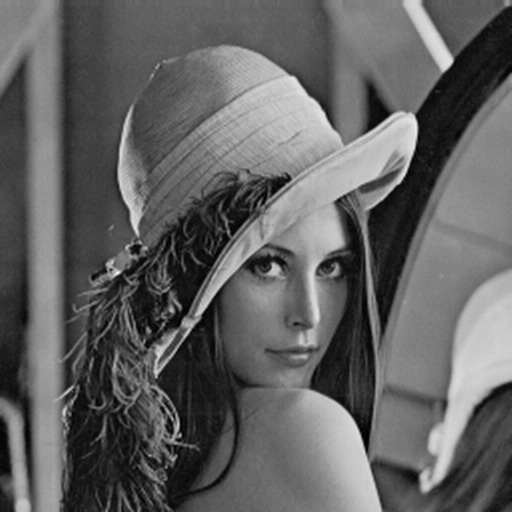}
    }
   \subfigure[$2^9\times2^9$]{
\includegraphics[width=0.9in]{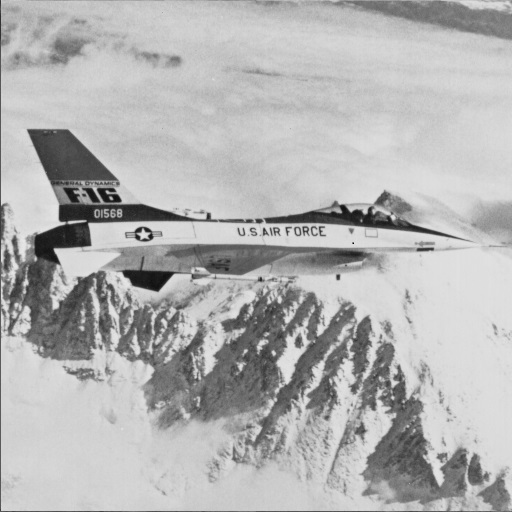}
   }
   \subfigure[2$^9\times2^9$]{
	\includegraphics[width=0.9in]{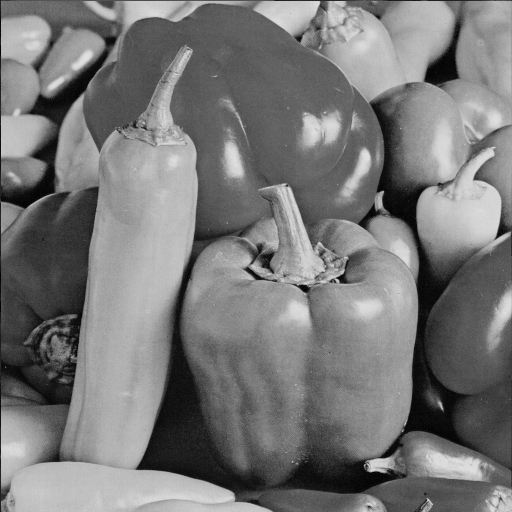}
    }
    \subfigure[$2^9\times2^9$]{
	\includegraphics[width=0.9in]{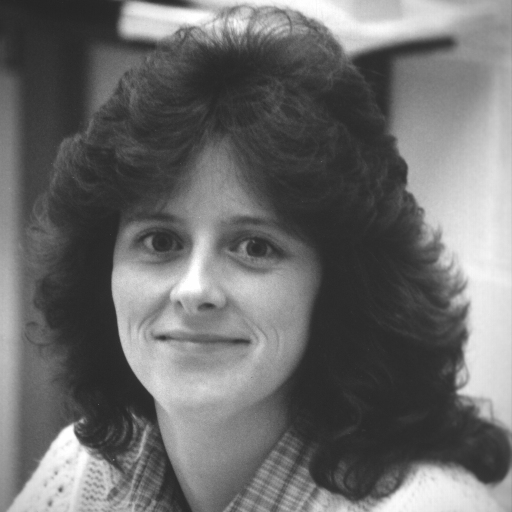}
    }
    \subfigure[$2^9\times2^9$]{
	\includegraphics[width=0.9in]{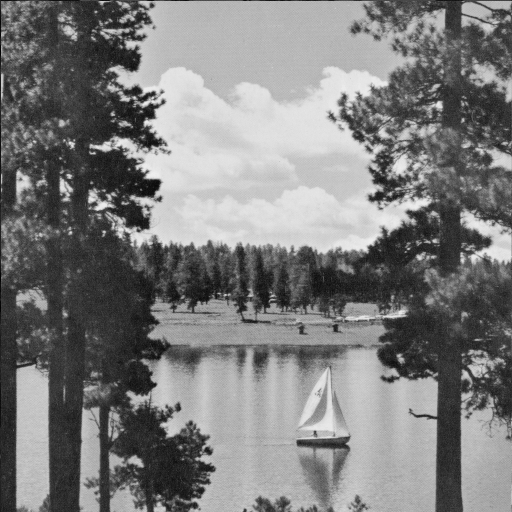}
    }
    \subfigure[$2^9\times2^9$]{
	\includegraphics[width=0.9in]{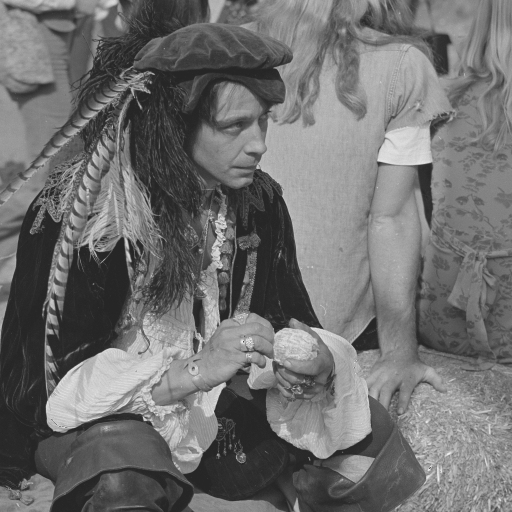}
    }
    \caption{Six images with $2^9\times2^9$-sized.}
    \label{figure15}
\end{figure}

\begin{figure}[htbp]
\centering
    \subfigure[N:$2^{11}\times2^{11}$]{
    	\includegraphics[width=0.9in]{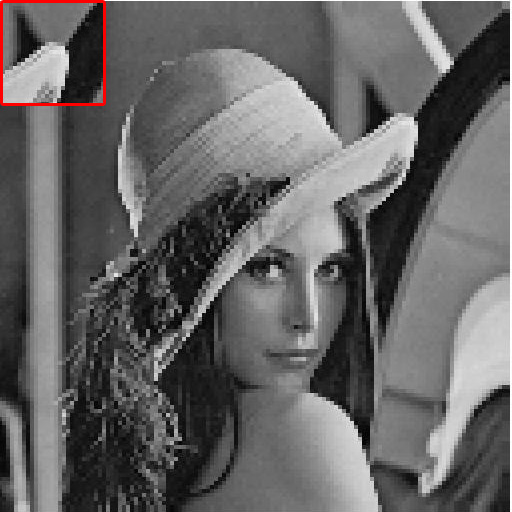}
    }
    \subfigure[N:$2^{11}\times2^{11}$]{
	\includegraphics[width=0.9in]{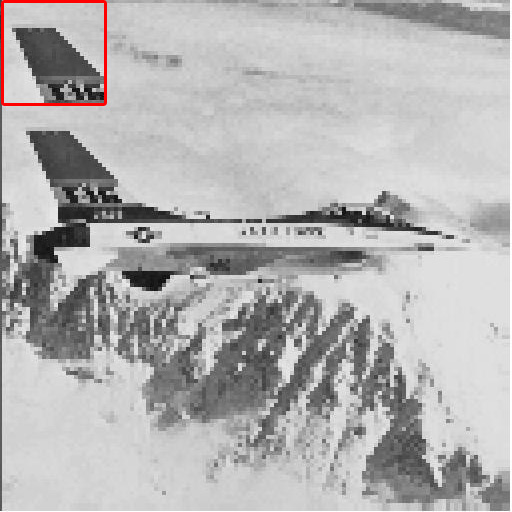}
    }
    \subfigure[N:$2^{11}\times2^{11}$]{
	\includegraphics[width=0.9in]{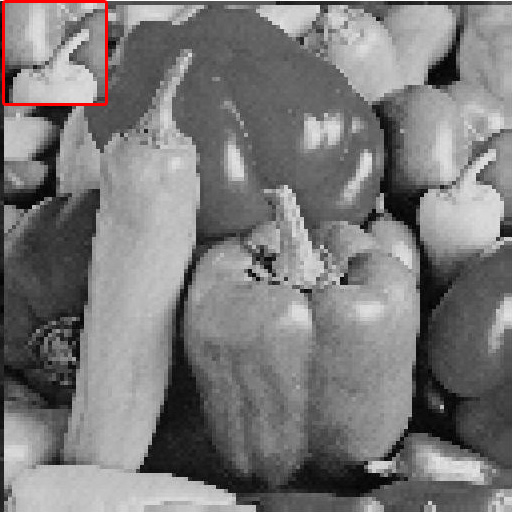}
    }
    \subfigure[N:$2^{11}\times2^{11}$]{
	\includegraphics[width=0.9in]{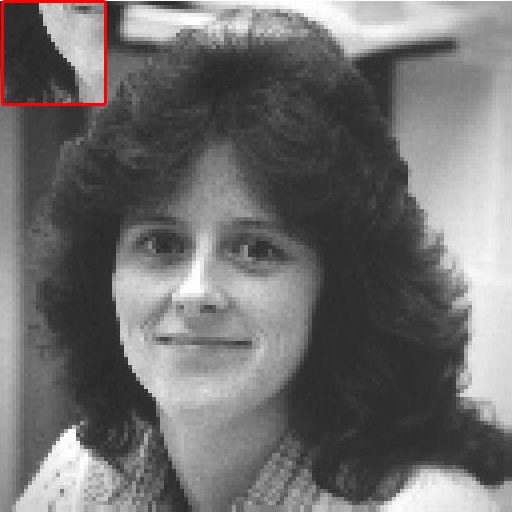}
    }
    \subfigure[N:$2^{11}\times2^{11}$]{
	\includegraphics[width=0.9in]{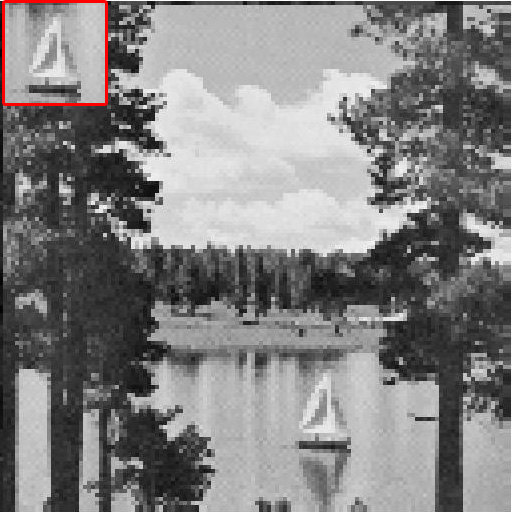}
    }
    \subfigure[N:$2^{11}\times2^{11}$]{
	\includegraphics[width=0.9in]{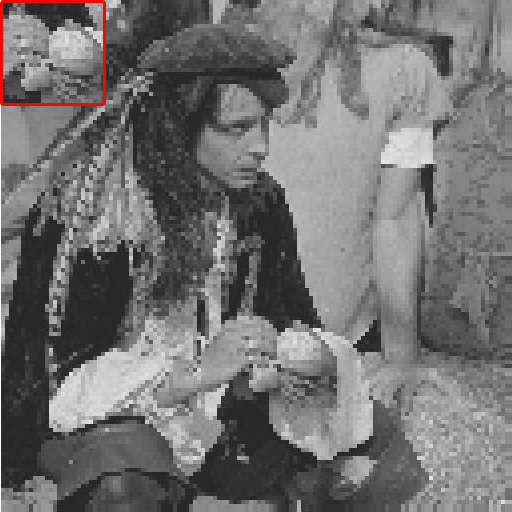}
    }
    \caption{Up-scaled images based on nearest neighbor with scaling ratio $2^2\times2^2$.}
     \label{figure16}
\end{figure}

\begin{figure}[htbp]
\centering
    \subfigure[B:$2^{11}\times2^{11}$]{
    	\includegraphics[width=0.9in]{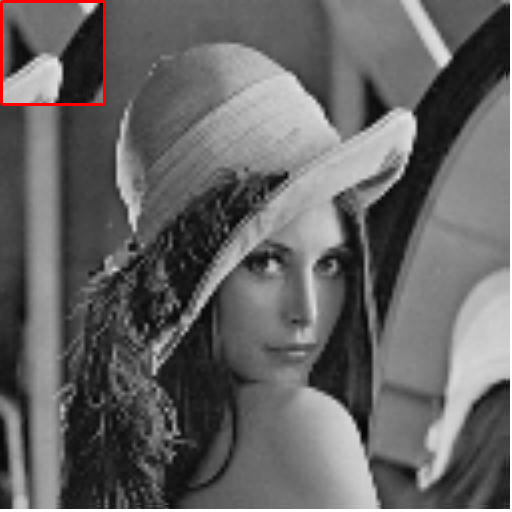}
    }
    \subfigure[B:$2^{11}\times2^{11}$]{
	\includegraphics[width=0.9in]{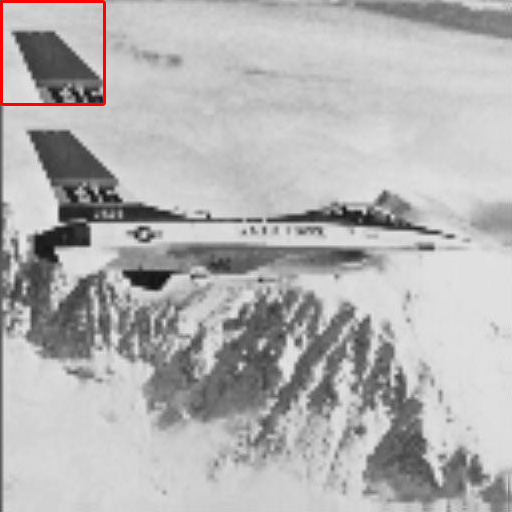}
    }
    \subfigure[B:$2^{11}\times2^{11}$]{
	\includegraphics[width=0.9in]{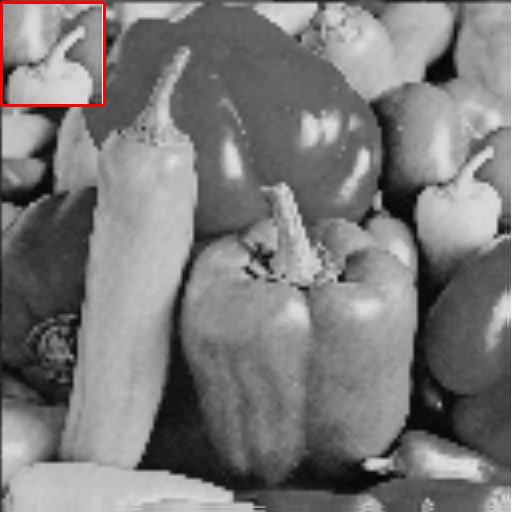}
    }
    \subfigure[B:$2^{11}\times2^{11}$]{
	\includegraphics[width=0.9in]{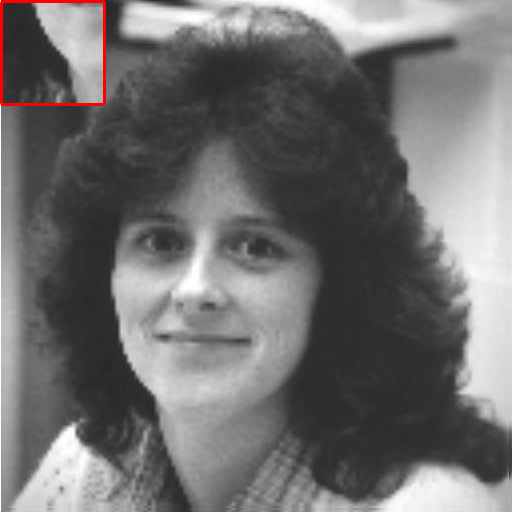}
    }
    \subfigure[B:$2^{11}\times2^{11}$]{
	\includegraphics[width=0.9in]{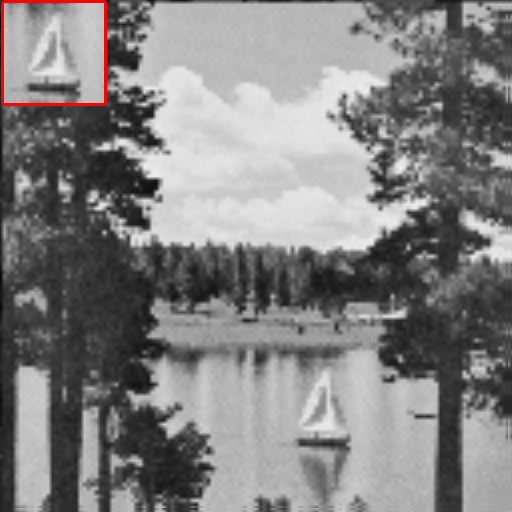}
    }
    \subfigure[B:$2^{11}\times2^{11}$]{
	\includegraphics[width=0.9in]{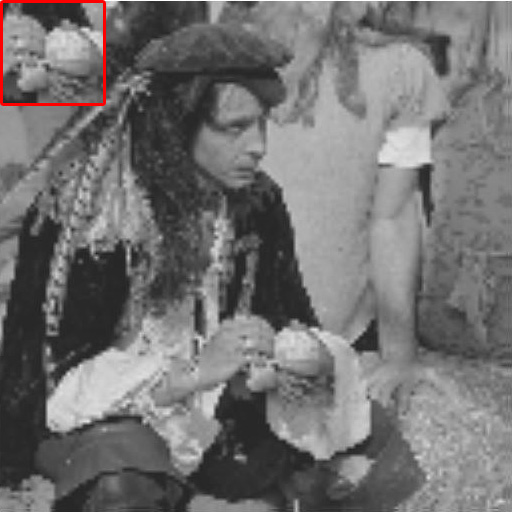}
    }
    \caption{Up-scaled images based on proposed bilinear with scaling ratio $2^2\times2^2$.}
    \label{figure17}
\end{figure}

In this section, we present the results of running our bilinear interpolation algorithm on FRQI images.  Our performance evaluation is built on simulation-based implementations on Matlab. We have taken six $2^9\times2^9$ test grayscale images  presented in Fig. \ref{figure15}. Figure \ref{figure16} presents up-scaled versions of those images presented in Fig. \ref{figure15} using the nearest neighbor scheme with scaling ratio ($2^1\times2^1$). Furthermore,  results produced by using our bilinear approach on the same set of original images (Fig. \ref{figure15}) are presented in Fig. \ref{figure17}. To assess the performance of these two different schemes, we use the peak signal-to-noise ratio (PSNR) \cite{li2018bilinear}\cite{iliyasu2012watermarking}\cite{li2014multidimensional} and the structural similarity (SSIM) metrics \cite {Hore2010image} for evaluation the up-scaled images and original images.

PSNR is the proportion between the maximum-possible power of an image and the power of corrupting noise that affects the fidelity of its representation. The higher the PSNR value, the better the image quality. Meanwhile, the SSIM index is a well-known quality metric used to measure the similarity between two images \cite{salazar-colores2019fast}, which is a decimal value between -1 and 1. SSIM = 1 only when two images with identical sets of data are compared. In the simulation experiments, the two parameters, i.e., the Gaussian kernel of standard deviation and the size of the window is set as the commonly used values, i.e., $\sigma =1.5$ and $\varpi=11\times 11$ \cite{wang2004image}. As stated above, the PSNR and SSIM results by different methods are listed in Table \ref{tbl:table1}.

\begin{table*}[h]
\caption{PSNR(dB) and SSIM results of up-scaled images obtained by two different methods.}
\centering
\begin{tabular}{lllllllllllll}
\toprule
\multirow{2}{*}{Scheme} & \multicolumn {2}{c}{(a)} & \multicolumn{2}{c}{(b)} & \multicolumn{2}{c}{(c)} & \multicolumn{2}{c}{(d)} & \multicolumn{2}{c}{(e)} &\multicolumn{2}{c}{(f)}\\
\cline{2-13}\\
& PSNR & SSIM & PSNR & SSIM & PSNR & SSIM & PSNR & SSIM & PSNR & SSIM & PSNR & SSIM\\
  \midrule
  N $2^1\times 2^1$ &  29.49& 0.75  & 26.37& 0.62  &26.67 &0.53&33.73&0.65&24.57&0.54&25.78&0.60\\
  \rowcolor{gray}B $2^1\times 2^1$  & 37.09 & 0.87 & 30.12 & 0.75 & 31.16 &0.63&38.44&0.76&28.86&0.66&29.83&0.73\\
  N $2^2\times 2^2$ & 22.67 & 0.34 & 21.69 & 0.26  &21.42&0.28&27.67&0.37&19.48&0.25&20.65&0.26\\
  \rowcolor{gray}B $2^2\times 2^2$  & 29.16 & 0.60 & 25.19 & 0.45 & 26.64&0.45&33.97&0.57&23.87&0.43&25.12&0.45\\
  \bottomrule
  \end{tabular}
  \label{tbl:table1}
\end{table*}

PSNR and SSIM index metrics for nearest neighbor (N)  and bilinear (B) techniques are presented in Table \ref{tbl:table1}. Experimental results depict that the proposed scheme produces better PSNR and SSIM results in comparison with the scheme proposed in \cite{sang2016quantum}. Additionally and as shown in Fig. \ref{figure16}, the up-scaled result of simulation experiment by using nearest neighbor interpolation algorithm have significant distortion and exhibit both mosaic and  tooth phenomena (shown in the red boxs). Figure \ref{figure17} shows the results using the bilinear interpolation scheme presented in this paper.

Results presented in Table \ref{tbl:table1} show the robustness of our method as both PSNR and SSIM index values for bilinearly scaled images are grater than corresponding nearest neighbor values. Furthermore, visual results for scaled images using our bilinear interpolation approach are of high quality.

\section{Conclusions}
\label{sec5}

In this paper, we have presented a  bilinear interpolation strategy to realize image up-scaling and down-scaling on FRQI quantum images. We have presented the quantum circuits that allow us to implement our approach in the quantum gate model and we have analyzed the circuit complexity of our proposal. Furthermore, the results of our experiments indicate that our bilinear interpolation scheme is robust and it has advantages in terms of PSNR and SSIM metrics compared with the nearest neighbor algorithm.

\section*{Acknowledgment}

This work was supported by Tecnologico de Monterrey, Escuela de Ingeneria y Ciencias and CONACyT (SNI No. 41594).

\bibliographystyle{ieeetr}      % mathematics and physical sciences
\bibliography{document}   % name your BibTeX data base
\end{spacing}
\end{document}